\documentclass[a4paper]{leccart}
\usepackage{epsfig}
\let\TT\textsuperscript
\title{Readout Electronics Tests and Integration of the ATLAS Semiconductor Tracker}
\author{Vasiliki A.~Mitsou\TT{a}\TT{b} for the ATLAS SCT Collaboration\\[14pt]
\TT{a} Instituto de F\'{i}sica Corpuscular (IFIC), CSIC -- Universitat de
Val\`{e}ncia, \\
Edificio Institutos de Investigaci\'{o}n, P.O.
Box 22085, E-46071 Valencia, Spain\\
\TT{b} CERN, PH-ATT Department, CH-1211 Geneva 23, Switzerland\\
\texttt{vasiliki.mitsou@cern.ch}}

\begin{document}
\maketitle

\begin{multicols}{2}
\begin{abstract}
The SemiConductor Tracker (SCT) together with the Pixel detector and the
Transition Radiation Tracker (TRT) form the central tracking system of the
ATLAS experiment at the LHC. It consists of single-sided microstrip
silicon sensors, which are read out via binary ASICs based on the DMILL
technology, and the data are transmitted via radiation-hard optical
fibres. After an overview of the SCT detector layout and readout system,
the final-stage assembly of large-scale structures and the integration
with the TRT is presented. The focus is on the electrical performance of
the overall SCT detector system through the different integration stages,
including the detector control and data acquisition system.
\end{abstract}

\section{Introduction}

The ATLAS detector \cite{atlas}, one of the two general-purpose
experiments of the Large Hadron Collider (LHC), has entered into the final
stages of installation at CERN. The LHC, a proton-proton collider with a
14-TeV centre-of-mass energy and a design luminosity of
\mbox{$10^{34}~{\rm cm^{-2}s^{-1}}$,} is expected to deliver the first
proton beam by the end of 2007. The ATLAS central tracker (Inner Detector,
ID) \cite{id} combines the silicon detector technology (pixels and
micro-strips) in the innermost part with a straw drift detector with
transition radiation detection capabilities (Transition Radiation Tracker,
TRT) in the outside, operating in a 2-T superconducting solenoid.

\begin{figure}
    \centering
    \epsfig{file=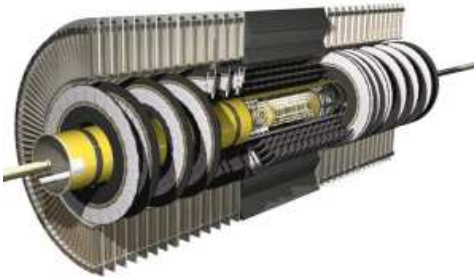,width=0.7\linewidth,clip=}
    \caption{Layout of the ATLAS Inner Detector: it comprises the Transition
    Radiation Detector, the Semiconductor Tracker and the Pixel system
    from the outer to the inner radii, respectively.} \label{fig:ID}
\end{figure}
\vspace*{-0.45cm}

The microstrip detector (Semiconductor Tracker, SCT), as shown in
Fig.~\ref{fig:ID}, forms the middle layer of the ID between the Pixel
detector and the TRT. The SCT system \cite{id,sct} comprises a barrel made
of four nested cylinders and two end-caps of nine disks each. The barrel
layers carry 2112 detector units {\em (modules)} altogether, while a total
of 1976 end-cap modules are mounted on the disks. The whole SCT occupies a
cylinder of 5.6~m in length and 56~cm in radius with the innermost layer
at a radius of 27~cm.

The silicon modules \cite{barrelmodule,endcapmodule} consist of one or two
pairs of single-sided p-\emph{in}-n microstrip sensors glued back-to-back
at a 40-mrad stereo angle to provide two-dimensional track reconstruction.
The $285$-$\mu{\rm m}$ thick sensors \cite{sensor} have 768 AC-coupled
strips with an $80~\mu{\rm m}$ pitch for the barrel and a $57-94~\mu{\rm
m}$ pitch for the end-cap modules. Between the sensor pairs there is a
highly thermally conductive baseboard. Barrel modules follow one common
design, while for the forward ones four different types exist according to
their position in the detector.

\section{The SCT readout system} \label{sec:twocol}

The readout of the module is based on 12~ABCD3TA ASICs manufactured in the
radiation-hard DMILL process mounted on a copper/kapton hybrid
\cite{hybrid}. The ABCD3TA chip \cite{ABCD} features a 128-channel analog
front end consisting of amplifiers and comparators and a digital readout
circuit operating at a frequency of 40.08~MHz. This ASIC utilises the
binary scheme where the signals from the silicon detector are amplified,
compared to a threshold and only the result of the comparison enters the
input register and a 132-cell deep pipeline, awaiting a level-1 trigger
accept signal. It implements a redundancy mechanism that redirects the
output and the control signals, so that a failing chip can be bypassed. To
reduce the channel-to-channel threshold variation, in particular after
irradiation, the ABCD3TA features an individual threshold correction in
each channel with a 4-bit digital-to-analog converter \emph{(TrimDAC)}
with four selectable ranges. In addition, a calibration circuitry is
implemented in the chip providing an injection charge in the range
$\rm0.5-10~fC$. By injecting various known charges and performing
threshold scans, the analogue properties of each channel can be
determined, such as the gain, the offset and the noise.

The clock and command signals as well as the data are transmitted from and
to the off-detector electronics through optical links \cite{optical}. On
the detector side, the DORIC\footnote{Digital Optical Receiver Integrated
Circuit.} and VDC\footnote{VCSEL (Vertical Cavity Surface Emitting Laser)
Driver Chip.} are utilised for receiving the optical clock and control
signal (one link) and for data transmission (two links), respectively.
Therefore, three optical fibres are connected to each module, terminated
by an opto-package consisting of Si \emph{p-i-n} diodes and VCSELs mounted
on the Back-Of-Crate (BOC) card. The latter serves as an interface between
the optical signals and the off-detector electronics in the Read-Out
Driver (ROD). Each ROD controls and monitors 48~SCT modules.

The LHC operating conditions demand challenging electrical performance
specifications for the SCT modules and the limitations mainly concern
acceptable noise occupancy level, tracking efficiency, timing and power
consumption. These requirements reflect on the design of the readout
system, as well as on the quality assurance/control strategy followed
throughout the detector construction. To this respect, a series of
electrical tests were performed during the various stages of the detector
assembly; from module production \cite{endcap_prod}, to macro-assembly
\cite{georg}, at reception at CERN \cite{mikulec}, and eventually after
the final integration with TRT. These repetitive tests are necessary to
ensure that the module/system performance does not change after each
stage, to finalise the corresponding data acquisition software, and to
learn how to recover potential errors/problems. The final stages of the
SCT assembly and testing at CERN were carried out in the ATLAS SR1
clean-room, equipped with a system capable of characterising up to one
million readout channels simultaneously. Electronics tests results such as
electrical connections checks, noise and gain measurements, as well as
temperature and leakage current measurements were given particular
attention. The outcome of these electronics tests, essential also for the
validation of the grounding, shielding and cooling system, are discussed
in the following sections.

\section{SCT assembly and integration}

\begin{figure}
    \centering
    \epsfig{file=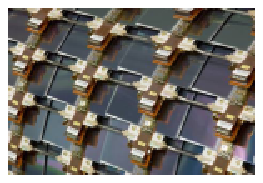,width=0.41\linewidth,clip=} \hfill
    \epsfig{file=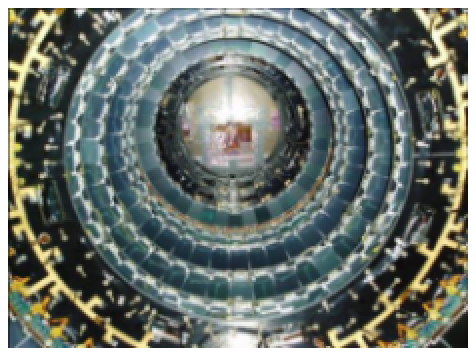,width=0.55\linewidth,clip=}
    \caption{Left: detail from a barrel outer surface; silicon sensors
    and hybrids are visible. Right: beam's eye view of end-cap~C.}
    \label{fig:detail}
\end{figure}

After the module production, being distributed over several sites in
Australia, Europe, Japan and the USA, the macro-assembly\footnote{Mounting
of modules onto disks (end-cap) or cylinders (barrel) and installation of
the respective services.} took place in three laboratories, in the UK and
the Netherlands. The four barrel layers were assembled at the University
of Oxford employing two specially designed robots for the mounting of
modules onto cylindrical support structures. A detail of a barrel layer
showing the overlapping modules is given in Fig.~\ref{fig:detail} (left).
The on-detector services, being fitted during macro-assembly, include
thin-wall Cu-Ni pipes and module cooling blocks, which remove heat through
an evaporative $\rm C_3F_8$ cooling system. The modules are individually
supplied with LV and HV power through Al- or Cu-\emph{on}-Kapton low mass
tapes.

\vspace*{-0.4cm}

The electrical performance of whole barrels was duly tested in Oxford for
barrels 3, 4 and~6\footnote{The four SCT barrels are numbered from 3 to 6
starting from the innermost one, \emph{i.e.}\ B3--B6, whereas B0--B2
denote the pixel layers.} and at CERN for barrels 3 and~5. All digital and
analog functions were examined by following standard measuring procedures
\cite{endcap_prod}. The tests showed stable operation for a simultaneous
readout of up to one million channels in terms of thermal operation, data
acquisition and detector control. More than 99.7\% of all barrel channels
are fully operational as evident from Table~\ref{tab:barrel}, where
detailed information for dead and noisy channels for each barrel is given.
These figures are in agreement with the ones obtained during module
production. The average module noise was stable at $4.5\times10^{-5}$,
well below the design specification of $5\times10^{-4}$. Concerning the
detector bias, the leakage current drawn by the sensors at
$\rm\sim15^{\circ}C$ was much lower than $\rm1~\mu A$ at the nominal bias
value of 150~V, \emph{i.e.}\ within the specifications.

\vspace*{-0.3cm}
\begin{table}
\caption{Channel defects breakdown for the barrel SCT as measured after
macro-assembly, in Oxford and at CERN.} \label{tab:barrel}
\renewcommand{\arraystretch}{1.2}
\centering
\begin{tabularx}{0.9277\linewidth}{|m{0.7cm}|m{1.2cm}|m{0.65cm}|m{0.65cm}|m{0.65cm}|m{1.8cm}|}\hline

Barrel & Total nr.\ channels & Dead & Noisy & Other & Total defects \\
\hline
B3     &    589\,824 &  357 &  460 &  666 & 1483 (0.25\%)      \\
B4     &    737\,280 &  245 &  242 &  354 &  841 (0.11\%)      \\
B5     &    884\,736 &  770 &  492 &  556 & 1818 (0.21\%)      \\
B6     & 1\,032\,192 & 2513 & 1936 & 1271 & 5720 (0.55\%)      \\ \hline
Total  & 3\,244\,032 & 3885 & 3130 & 2847 & 9862 (0.30\%)      \\ \hline
\end{tabularx}
\end{table}

After completing the individual testing of the barrels at CERN SR1 room,
the four layers were eventually integrated into one barrel. This
operation, completed within a period of three months, was carried out
step-by-step with one layer being inserted into the structure each time
from the largest down to the smallest one. During this operation, the SCT
services of the \emph{inner} layer were transferred onto a horizontal
service support structure, whereas the ones of the \emph{outer} layer were
unfolded radially at the ends.

\begin{figure}
    \centering
    \epsfig{file=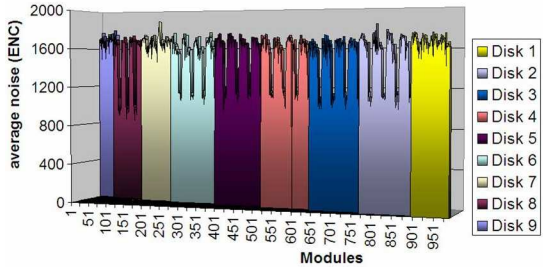,width=0.8\linewidth,clip=}
    \caption{ENC noise per module for all disks of end-cap A as measured
    at CERN. The modulation observed in the plot is due to the
    different noise levels of the four types of end-cap modules
    \cite{endcapmodule}; $\rm\sim1450-1550~e^-$ for the outer and long middle
    modules and $\rm\sim900-1100~e^-$ for the inner and short middle
    ones.}
\label{fig:ECnoise}
\end{figure}

The two end-caps, on the other hand, were brought together at the
University of Liverpool (End-Cap~C, \emph{EC-C}) and at NIKHEF (End-Cap~A,
\emph{EC-A}). The modules were manually mounted onto disks, fully
characterised in a test-box and finally the disks were installed inside
carbon fibre cylinders. A photograph of the fully assembled EC-C as seen
from inside is shown in Fig.~\ref{fig:detail} (right); all disks are
clearly visible. Both end-caps have been transferred to CERN ---first EC-C
and then EC-A--- and they have passed successfully the reception tests,
which include visual inspection, disk alignment measurements, examination
of the cooling circuits and comprehensive electronics tests. The EC-A was
tested while cooled down to the nominal temperature of $\rm-7^{\circ}C$ in
contrast to EC-C which was tested warm. The measured
ENC\footnote{Equivalent Noise Charge defined as the input charge giving
signal equal to effective output noise (expressed in electrons).} noise
for each module of EC-A is shown in Fig.~\ref{fig:ECnoise}. These values
are comparable to the ones measured during module assembly
\cite{endcap_prod} and after macro-assembly at NIKHEF. The fraction of
dead channels was also found to be at the same level as previously
measured and around 0.2\%.

Many of the activities performed on the SCT large structures was devoted
to connection and manipulation of services, \emph{i.e.}\ the connection of
the evaporative cooling plants, the final power supplies and readout
electronics. A prototype of the final Data Acquisition (DAQ) system,
consisted of RODs and BOCs, and the Detector Control System (DCS) was
engaged. The DCS provides high and low voltage for the sensors and ABCD3TA
chips, voltage for the DORIC and VDC ASICs and monitors the temperature on
the modules. Besides monitoring and controlling various parameters of
these power supplies, the DCS includes a hard-wired interlock system which
automatically switches off the power supply to certain groups of modules
in the event of an over-temperature. In addition, the DCS monitors
humidity and temperature sensors mounted on the outlets of each cooling
loop and records the corresponding data.

\vspace*{-0.4cm}

Each of the three SCT blocks, the barrel and the end-caps, are enfolded in
cylindrical structures, the Outer and the Inner Thermal Enclosures (OTE
and ITE, respectively). These foam-based layers, covered by aluminised
Kapton (OTE) or carbon-fibre-reinforced-plastic (ITE), provide gas
tightness,\footnote{The SCT will be operated in a $\rm N_2$ environment at
$\rm-7^{\circ}C$, whereas the TRT will be embedded in $\rm CO_2$ at a
temperature of $\rm\sim20^{\circ}C$.} thermal isolation and Faraday
shielding. The gas envelopes are complemented by flat panels fitted on the
ends. A quite large fraction of testing time was given to gas tightness
measurements, identification and sealing of leaks. After several
iterations, the leak rate was reduced to an acceptable level for the
barrel and EC-C (EC-A OTE has not been fitted yet).

\section{Barrel SCT-TRT integration}

\begin{figure}
    \centering
    \epsfig{file=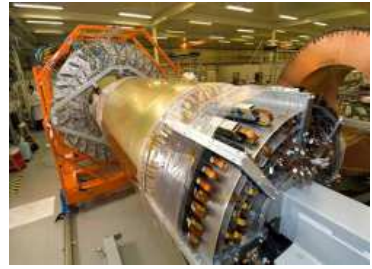,width=0.54\linewidth,clip=}
    \caption{Insertion of barrel SCT into barrel TRT; the OTE is
    visible surrounding the barrel SCT, as well as the services on the ISSS
    (foreground).}\label{fig:SCTintoTRT}
\end{figure}

The integration and commissioning of the barrel SCT with the respective
TRT \cite{trt_status} is almost complete. It started in February~2006 with
the insertion of one detector into the other, shown in
Fig.~\ref{fig:SCTintoTRT}, using a rail system and a cantilever stand.
During this operation, the SCT services were transferred onto the
Insertion Service Support Structure (ISSS), fixed onto the SCT cradle
extensions. The TRT, installed in the inner detector trolley, was finally
slid over the SCT.

\begin{figure}
\centering
\begin{minipage}[c]{0.48\linewidth}
  \centering\epsfig{file=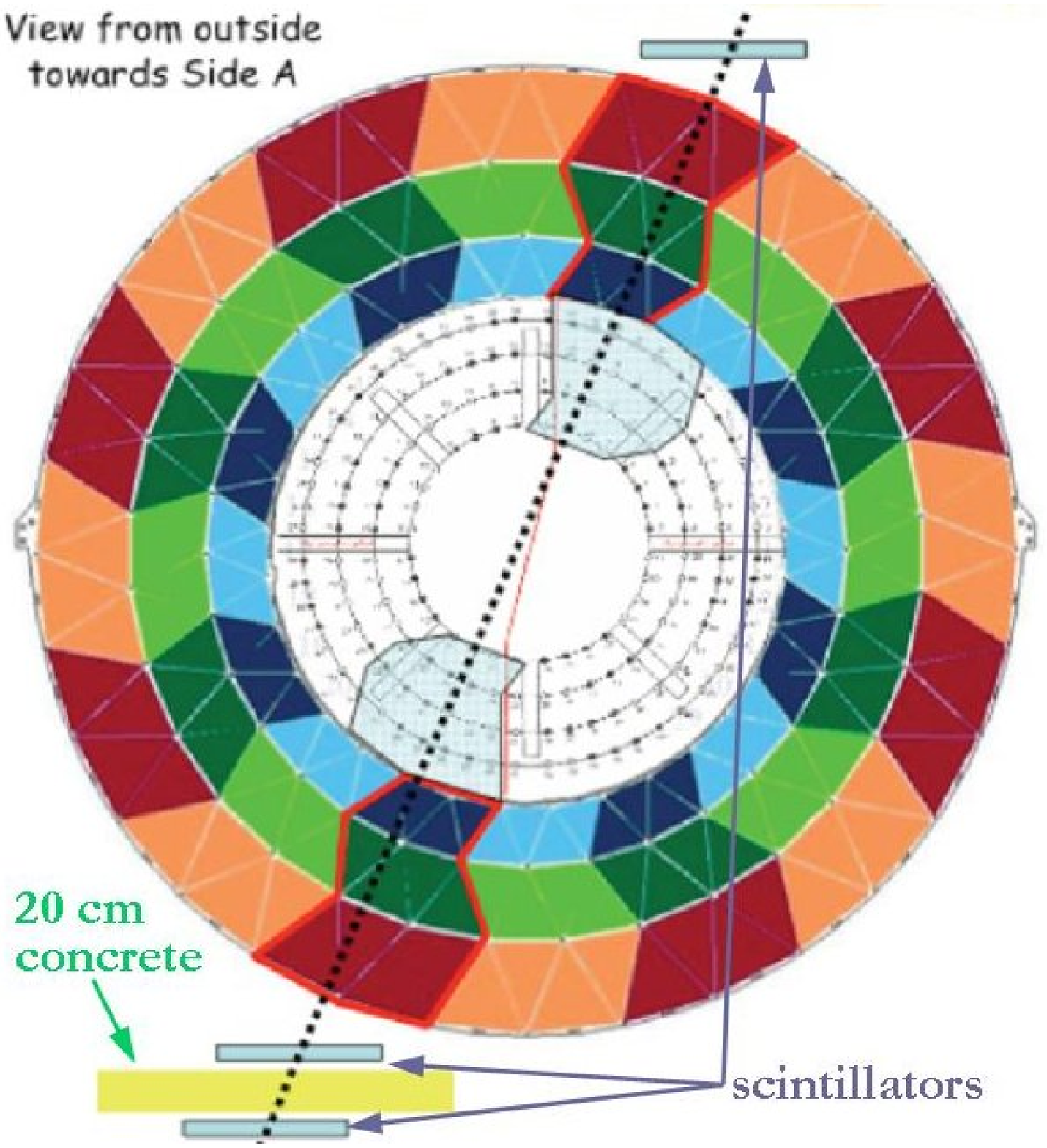,width=\linewidth,clip=}
\end{minipage} \hspace{0.03\linewidth} 
\begin{minipage}[c]{0.4\linewidth}
  \vspace*{-2mm}
  \epsfig{file=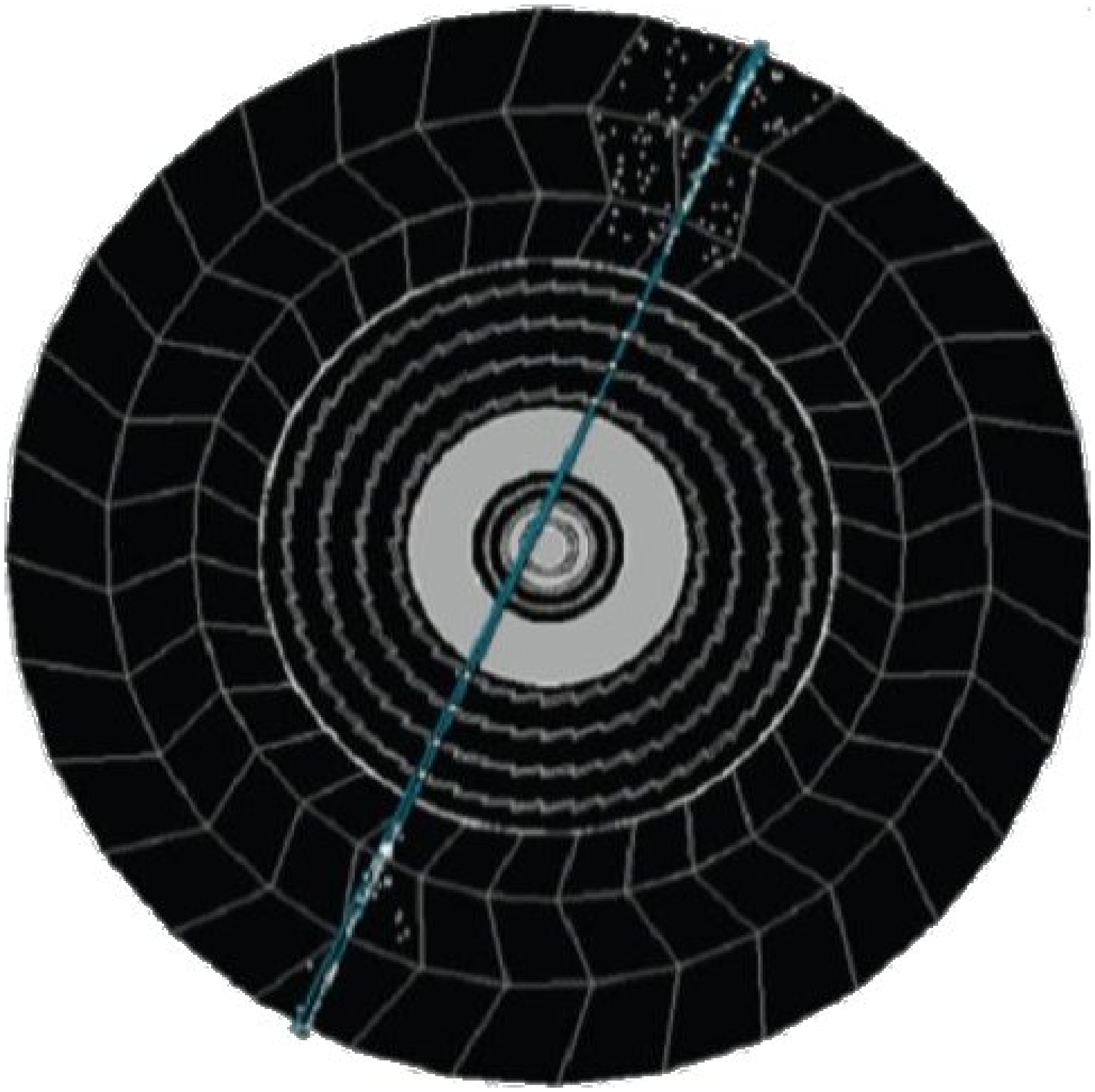,width=\linewidth,clip=}
\end{minipage}
 \caption{Left: Transverse schematic view of the setup for the cosmic run.
    The actually read out parts are highlighted (half of the bottom TRT section
    was not read out). Three scintillator counters were used for trigger. Right:
    reconstructed cosmic track in the SCT and the TRT.}
    \label{fig:cosmics}
\end{figure}

A series of combined tests followed the integration of the barrel ID,
covering a wide spectrum of operational and detector performance related
aspects \cite{cosmics}. During these tests in SR1, one eighth of the TRT
and one quarter of the SCT were equipped with the complete readout chain,
in a top-bottom layout as shown in the left-hand side of
Fig.~\ref{fig:cosmics}. As far as the SCT is concerned, 468 out of the
2112 modules were read out using 12~RODs and one TIM.\footnote{TTC
(Timing, Trigger \& Control) Interface Module.} In the TRT, on the other
hand, about 10\,000 channels were examined with nine RODs and three TTCs.
Three scintillator counters were also installed (see
Fig.~\ref{fig:cosmics}, left) to provide an external trigger from cosmic
rays to both detector systems. A typical cosmic-ray track is shown on the
right-hand side of Fig.~\ref{fig:cosmics} as reconstructed by a combined
SCT \& TRT tracking algorithm. The measurement results that follow were
obtained with this \emph{cosmic} setup.

\vspace*{-0.5cm}

\vspace*{-0.4cm}

A typical electrical testing sequence begins by establishing communication
between the modules and the off-detector electronics system and by
optimising the optical links settings. After the digital tests,
\emph{i.e.}\ checking the redundancy links, the chip by-pass functionality
and the pipeline circuit, the analogue measurements follow. These include
measurements of the gain, the offset and the noise for each channel and
evaluation of the module noise occupancy. The noise is measured by
performing a threshold scan in the absence of charge injection. The slope
of the logarithm of the occupancy versus the square of the threshold is
approximately proportional to the ENC noise. Furthermore, the noise
occupancy at the 1-fC threshold level is obtained, with the exact 1-fC
point for each channel already been defined by the trimming. Several
distinct configurations were tried in order to assess the potential
dependence of the noise on those.

The noise stability was monitored throughout the measurements and only a
slight increase was observed for single barrels in comparison with the
macro-assembly values, as well as between the whole barrel and the
individual barrel layers. No pick-up noise was detected in the presence of
external heaters on the SCT thermal enclosure. After applying temperature
corrections, the ENC noise was found to be $\rm\sim40-50~e^{-}$ higher for
the barrel than after macro-assembly.

\begin{figure}
    \centering
    \epsfig{file=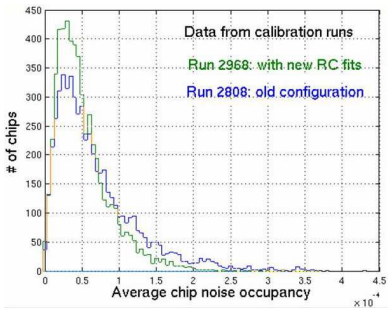,width=0.57\linewidth,clip=}
    \caption{Average noise occupancy for each chip for the cosmic test setup
    and for two different threshold configurations: the`old trim target'
    (blue), rendering a mean (r.m.s.) value of $6.8\times10^{-5}$
    ($5.5\times10^{-5}$), and the `new RC' (green), returning $5.1\times10^{-5}$
    ($3.7\times10^{-5}$), respectively. The measurements with RC configuration
    were taken after a few days with detector bias.}
    \label{fig:NObarrel}
\end{figure}

For the noise occupancy measurements, the TrimDAC thresholds for each
channel were initially set to the value obtained during module production,
leading however to a wide threshold variation. To remedy this problem the
so-called \emph{new Response Curve (RC)} configuration was introduced,
which included a $\sim10\%$ wafer-by-wafer correction to account for
variations in the ABCD3TA calibration capacitor. The latter configuration
is compared with the \emph{old trim target} uncorrected one in
Fig.~\ref{fig:NObarrel} with respect to the measured noise occupancy. With
the new configuration, the noise occupancy values are clearly less
scattered, returning an r.m.s.\ value of $3.7\times10^{-5}$, whereas the
old one gives an r.m.s.\ of $5.5\times10^{-5}$. The observed decrease in
the mean value with the RC threshold, on the other hand, is due to the
long period of detector bias preceding these measurements, rather than the
threshold configuration. This behaviour had previously been observed in
single barrel tests and is evident in calibration as well as physics-mode
runs. The RC configuration was used throughout the cosmic run.

\vspace*{-0.4cm}

The effect of high trigger frequency was studied by varying the pulser
rate in physics mode. For a trigger rate ranging from $\rm5~Hz$ up to
$\rm50~kHz$, no evidence of increase in noise occupancy was found.

The grounding scheme may be a potential factor contributing to noise
level. Therefore the effect of a change in the grounding was studied, by
measuring the noise occupancy with the power supply DC shorting cards
\emph{in} and \emph{out} when noth SCT and TRT are triggered from a pulser
at $\rm50~Hz$. No significant change in the noise occupancy was observed,
however the grounding scheme in SR1 is not the final one, which will not
be available before the detector is installed in its final position.

Noise can be evaluated by two methods: online with calibration scans and
triggers provided by the ROD and TIM and offline with physics mode runs
triggered by a pulser by applying the offline analysis. With either
approach practically the same results were obtained in terms of noise
occupancy.

\begin{figure}
    \centering
    \epsfig{file=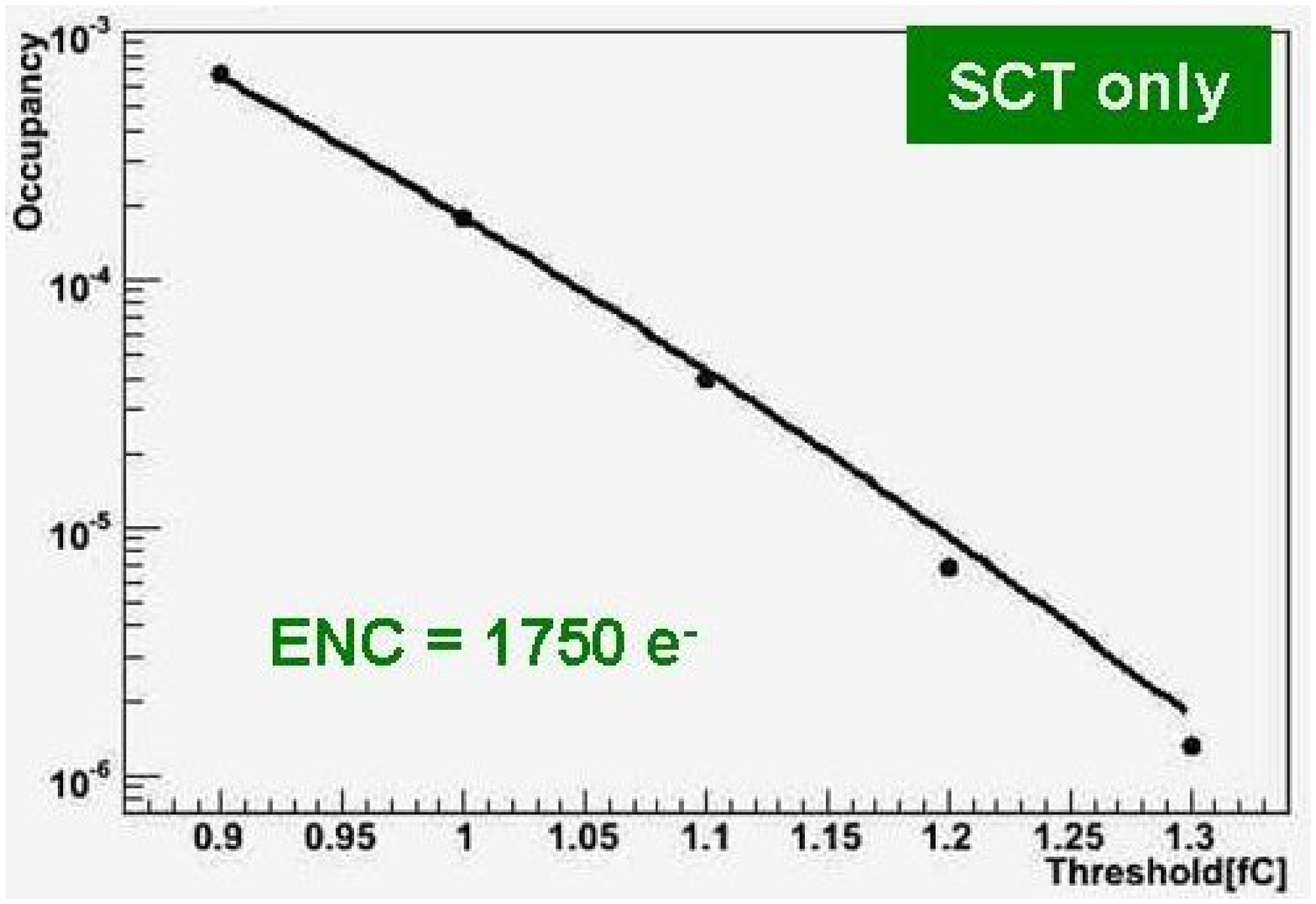,width=0.48\linewidth,clip=}\hfill
    \epsfig{file=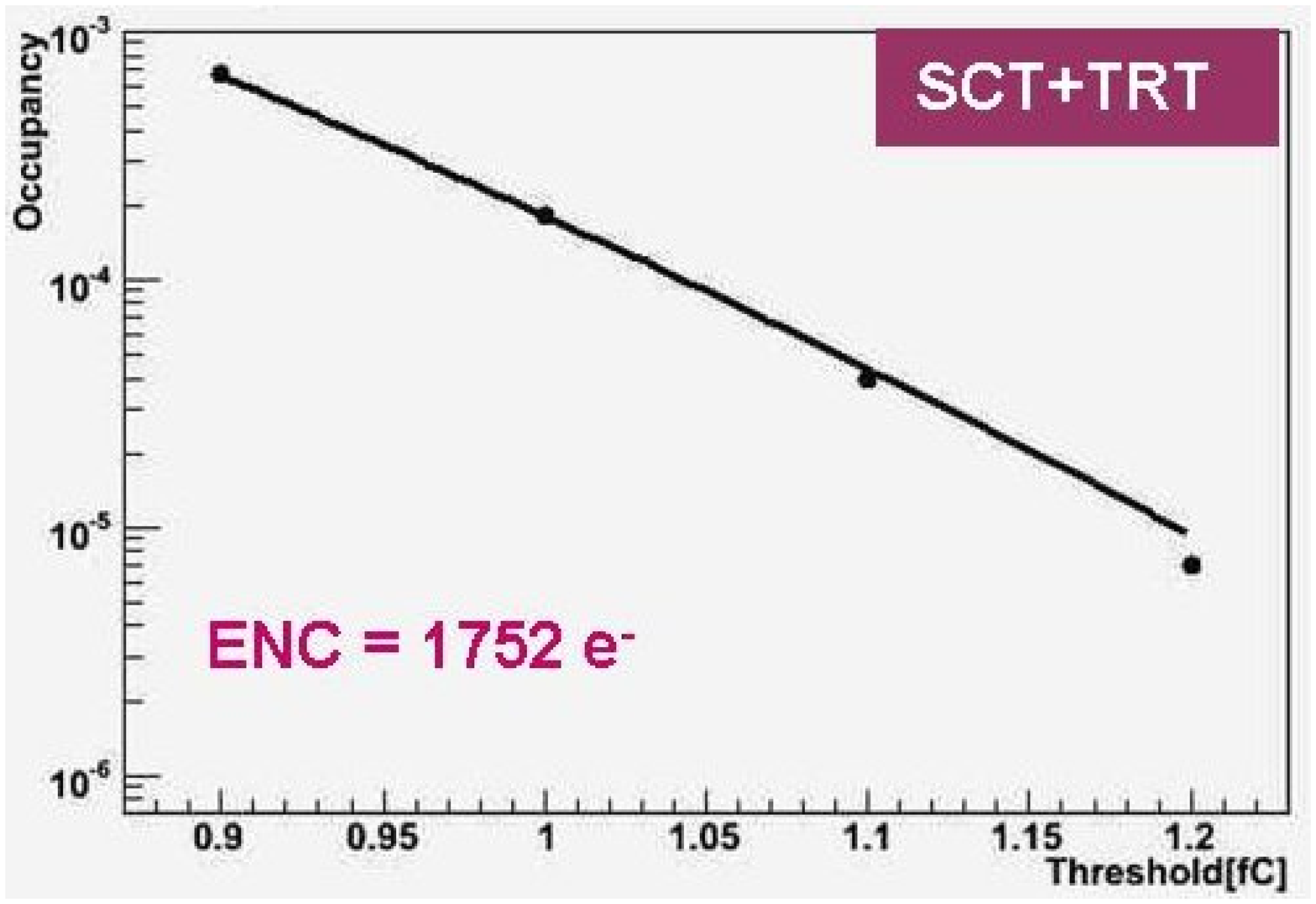,width=0.48\linewidth,clip=}
    \caption{Noise occupancy as a function of the threshold for SCT only
    (left panel) and for SCT and TRT simultaneous readout (right panel) when
    the \emph{test mode} (XXX, excluding 000) is selected for the data compression.
    The points represent measurements taken with threshold scans and the
    curves are analytically (through a complementary error function)
    derived from the ENC value indicated.}
    \label{fig:SCT}
\end{figure}

In Fig.~\ref{fig:SCT}, a comparison between the ENC noise recorded when
the SCT only is read out (left panel) and when the TRT is also operated
simultaneously (right panel). The ENC noise in both cases remains the same
and equal to $\rm\sim1750~e^{-}$, thus no electrical pick-up noise is
induced between the two detectors. A similar result is acquired when a
different data compression logic\footnote{In view of the foreseen 1\%
strip occupancy on any event, data compression is employed, in order to
reduce the number of bits of data read out of the chip for each event.
This logic examines the three bits of data, representing three beam
crossings centred on the level-1 trigger time, making up the hit pattern
for each channel. The state `X' denotes any bit value; 0 or 1.} than the
previously applied \emph{test mode} (XXX, excluding 000) is selected; for
the \emph{level mode} (X1X), the ENC noise is 1611~electrons for the SCT
only and 1610~electrons for SCT \& TRT together. Similar conclusions are
drawn for the TRT in the absence/presence of the SCT readout.

\vspace*{-0.4cm}

As far as common-mode noise is concerned, no evidence of such was found.
There was no increase in noise occupancy observed when using synchronous
triggers. No correlations have been observed neither between noise hits
within chips, nor between hits on different modules.

\begin{figure}
    \centering
    \epsfig{file=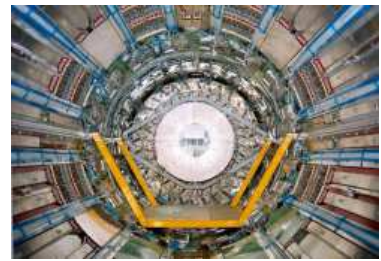,width=0.55\linewidth,clip=}
    \caption{The barrel ID installed in its final position inside the
    electromagnetic calorimeter cryostat bore.} \label{fig:barrel_pit}
\end{figure}

The ID barrel was transported from SR1 building and installed in the ATLAS
detector in August~2006. After being lowered into the ATLAS cavern with
only a few millimeters of clearance, the detector was successfully
inserted in the liquid argon calorimeter cryostat (see
Fig.~\ref{fig:barrel_pit}). The final stage of the barrel ID commissioning
is under way, involving the connection of cables and services, the
verification of full connectivity to the power supplies, readout and DAQ
systems, and the on-detector functionality checks of all detector modules
for SCT and TRT. This intervention will be carried out by implementing the
final grounding and shielding scheme for all $\sim2000$ SCT modules and
the complete TRT. It should validate the detector for autonomous operation
for ten years without further access.

\vspace*{-0.4cm}

\section{End-cap SCT-TRT integration}

\begin{figure}
    \centering
    \epsfig{file=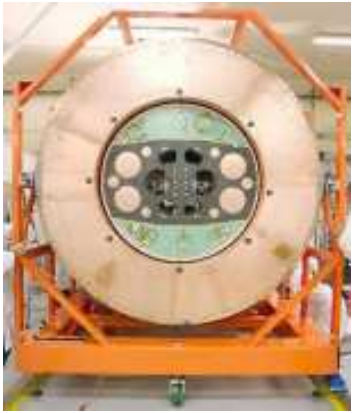,width=0.52\linewidth,clip=}
    \caption{Insertion of the SCT EC-C into the TRT EC-C.} \label{fig:IDC}
\end{figure}

The integration of the forward parts of the ID started in September 2006
with the insertion of the SCT EC-C into the TRT EC-C, shown in
Fig.~\ref{fig:IDC}. Before this operation, the functionality of an octant
of the EC-C was successfully tested inside the thermal enclosures. The
readout and power cables are currently being connected in preparation for
the two-month long combined tests in the SR1 clean-room. During those, one
quadrant of the SCT, \emph{i.e.}\ 247 modules, will be read out together
with an adjacent sector of the TRT corresponding to 1/16 of the total
end-cap (7680 straws/channels).

\vspace*{-0.4cm}

The ID EC-A, on the other hand, is going to follow more or less the same
integration and installation steps as EC-C. However, this process is
expected to proceed faster taking advantage of the experience gained
during the EC-C integration. The ID EC-A is expected to be integrated by
November. Both end-caps are scheduled to be ready for installation in the
ATLAS cavern by January -- February~2007.

\section{Conclusions}

ATLAS SCT is progressing well towards integration with the other parts of
the ID, installation in ATLAS and commissioning. Repeated tests in various
stages have demonstrated operational stability and good electrical
performance. The fraction of dead channels has been kept below $0.2\%$.
Particular attention was given to the electronics tests such as electrical
connections checks, noise and gain measurements, as well as temperature
and leakage current measurements. The outcome of these tests validated the
grounding, shielding and cooling system. As far as noise is concerned, no
remarkable change with respect to measurements during module production
and macro-assembly has been observed and no pick-up noise has been
detected while TRT is read out. Combined tests with cosmic rays allowed to
gain experience with the overall operational and running conditions (DAQ,
DCS, monitoring, \emph{etc}). The barrel ID (SCT \& TRT) has been
successfully installed in the ATLAS cavern inside the electromagnetic
calorimeter cryostat and the SCT end-caps integration with the TRT has
already started and is well under way. The innermost layer of the ID, the
Pixel detector, will be installed independently in 2007.

\section*{Acknowledgments}

The progress reported in this paper represents work performed across the
ATLAS SCT collaboration. I would like to thank all my colleagues whose
work is presented here and, in particular, Pepe Bernabeu for his useful
comments and suggestions.

\end{multicols}

\end{document}